# When Heating Isn't Cooling in Reverse: Nosé–Hoover Thermostat Fluctuations from Equilibrium Symmetry to Nonequilibrium Asymmetry

Hesam Arabzadeh[1†] and Brad Lee Holian[2‡]

[1]*Department of Chemistry, University of Missouri, Columbia, Missouri 65211-7600, USA*
[2]*Theoretical Division, Los Alamos National Laboratory, Los Alamos, New Mexico 87545, USA*

(‡Electronic mail: blhksh@gmail.com)
(†Electronic mail: hacr6@missouri.edu)

**ABSTRACT**

Recent laboratory experimental work has suggested that heating and cooling processes may exhibit intrinsic asymmetry, with heating occurring more efficiently than cooling. Two decades earlier, nonequilibrium molecular dynamics simulations addressed the topic of heating one side of a computational cell while cooling the other side by applying two different thermostats, producing a sinusoidal temperature profile. We revisit the theory underlying those computer calculations, and show how it accurately predicts the laboratory results. Recent realizations of a simple two-dimensional one-particle cell model give surprisingly relevant results, where two Nosé-Hoover thermostats are applied to the two directions ($x$ and $y$) at two temperatures, $T_x \geq T_y$. At equilibrium, the thermostatting rate variables $\xi_x$ and $\xi_y$ are identical, while under nonequilibrium temperature differences, the heating variable ($\xi_x$) is weaker than the cooling one ($\xi_y$), demonstrating the ratio of thermostat effort to thermal bias, just as predicted by theory: $\langle \xi_x \rangle / \langle \xi_y \rangle = -T_y/T_x$. We relate this to the negative rate of change in entropy of the nonequilibrium system that accompanies the contraction of phase space onto a fractal strange attractor of lower dimension. Histograms of the thermostat variables reveal the stark differences between equilibrium and nonequilibrium heat flow. At equilibrium, the cell-model $\xi$-distributions are both Gaussians centered at zero. We re-did much earlier many-body simulations of a Nosé-Hoover thermostatted fluid at equilibrium, which reported that the distribution was biased toward heating; we discovered that the prior work suffered from systematic integration error. In fact, we find that the distribution at equilibrium is a totally symmetric Gaussian for many-body systems, in agreement with the cell model. Under nonequilibrium conditions, when $T_x > T_y$, the simple 2D cell model clearly demonstrates the microscopic origin of heating–cooling asymmetry, thereby strongly confirming the results of real-world laboratory experiments.

## I. INTRODUCTION

*"C'est la dissymétrie qui crée le phénomène."* (*It is the asymmetry that creates the phenomenon.*) ─ Pierre Curie, *Journal de Physique* **3**, 393 (1894)

The laws of thermodynamics are generally assumed to treat heating and cooling on equal footing. Microscopic time-reversibility implies that energy can, in principle, be injected or removed with equivalent ease, subject only to fluctuations. Yet recent laboratory experiments have cast doubt on this intuitive symmetry: Ibáñez et al. [1] demonstrated that heating a system occurs more efficiently – faster and with less dissipation – than cooling the same system by the same amount.

The Ibáñez et al. heating and cooling experiments were conducted on colloidal particles (silica microspheres of diameter one micrometer) with charged surfaces, suspended in water, confined by a near-infrared laser trap, and subjected to a heat bath produced by a white-noise electric field. Particle Brownian motion was tracked by laser scattering for three temperature quench protocols: relaxation between either a higher temperature or lower to an intermediate target temperature, then the reverse procedure starting from equilibration at the target temperature, and finally, either heating or cooling between only two temperatures. In all cases, heating is seen to be more easily accomplished than cooling when a nonequilibrium temperature difference exists; otherwise, at equilibrium, there is no bias. This nonequilibrium asymmetry raises important questions: Is the asymmetry in heating and cooling a purely macroscopic effect? Or does it emerge naturally at the microscopic scale?

Molecular-dynamics (MD) computer simulations at the atomistic scale have used integral feedback to maintain constant temperature, employing the Nosé-Hoover (NH) thermostat [2]: the usual Newtonian equations of motion are augmented by a "frictional" term that is added to the momentum rate of change for each atom, namely, minus a thermostatting rate $\xi$ times the particle's momentum. The NH thermostat variable $\xi$ is an additional phase-space degree of freedom (beyond the atomic coordinates and momenta), which mimics energy exchange between the particles and an external heat bath. When $\xi > 0$, the bath takes energy out of all particles by slowing them down (frictional); when $\xi < 0$, the thermostat pumps energy into the particles, speeding them up (anti-frictional); at equilibrium $\xi$ fluctuates around zero. NH thermostatting at equilibrium makes the fluid behave as though it were sampling the NVT canonical ensemble ($N$ particles in volume $V$ and temperature $T$).

By Fourier's Law, heat flows irreversibly "downhill" from a hot temperature $T_1$ to a cold temperature $T_0$, where $T_1 > T_0$. We have just recently studied the two-dimensional single-particle cell model [3], where a wanderer particle moves in a periodic box with confinement provided by repulsive "mounds" at the four corners, and temperature in the two directions are orthogonally thermostatted: one direction hot ($x$) at temperature $T_1 = T_x$ and one cold ($y$) at $T_0 = T_y$. This cell model displayed the same asymmetry in the Lyapunov spectrum under nonequilibrium conditions as it does for the two thermostatting coefficients, $\xi_0$ and $\xi_1$, but no sign of asymmetry at all when $T_x = T_y$.

In 2002, a many-body nonequilibrium molecular-dynamics (NEMD) simulation of heat flow [4] was inspired by an imaginary thought experiment: a Maxwell Demon separates a hot reservoir at temperature $T^{(1)}$ (steady-state average is $T_1 > T_0$) from a cold sample at temperature $T^{(0)}$ (steady-state average is $T_0$) , using an invisible wall; sample and reservoir particles interact across the boundary at $x = 0$ by terms in the total potential energy, $\Phi$. Particles can move freely between the two sides and are subject to periodic boundary conditions (PBC). The Demon applies two Nosé-Hoover thermostats: for any sample particle at $x < 0$, the rate is $\xi_0$ , whose long-time average is positive, so as to extract heat from the cold sample; for any reservoir particle at $x > 0$, the rate is $\xi_1$, whose long-time average is negative, so as to pump heat into the hot reservoir. Let us assume that there are an equal number of particles, $N$ on average, in both the hot reservoir and cold sample. Of course, the temperature profile smoothes out to become approximately sinusoidal, which allows the resulting simulation, with decreasing temperature difference, to be analyzed by Navier-Stokes-Fourier hydrodynamics to yield the thermal conductivity at the Green-Kubo (equilibrium) limit. (Similarly, shear viscosity can be obtained from a system where one side has a fluid velocity in the y-direction imposed by a fluctuating acceleration, while the other has the opposite driving; in this case, only one NH thermostat keeps the temperature constant on both sides; it's the velocity profile that smoothes out to be nearly sinusoidal.)

This Demonic NEMD thought experiment allows us now to make some general arguments about the entropic origin of the asymmetry of nonequilibrium heating versus cooling. Focusing on one side or the other of the Maxwell Demon system, the volume $V$ contains $N$ atoms, characterized by the flow velocity of the center-of-mass (CM). Subtracting out that CM velocity results in what is called the "peculiar" (or "thermal") velocity for each particle.

Total mass, $M$: $M = \sum_{i=1}^{N} m_i$

($m_i$ = mass of particle $i$, $N$ particles)

CM (center of mass) $\mathbf{x}_0$: $M\mathbf{x}_0 = \sum_{i=1}^{N} m_i \mathbf{x}_i$

($\mathbf{x}_i$ = coordinate of particle $i$)

CM velocity $\dot{\mathbf{x}}_0$: $M\dot{\mathbf{x}}_0 = \sum_{i=1}^{N} m_i \dot{\mathbf{x}}_i$

($\dot{\mathbf{x}}_i$ = velocity of particle $i$)

Peculiar velocity of particle $i$: $\mathbf{u}_i = \dot{\mathbf{x}}_i - \dot{\mathbf{x}}_0$

($\mathbf{p}_i = m_i \mathbf{u}_i$ = momentum , $\sum_{i=1}^{N} m_i \mathbf{u}_i = 0 = \sum_{i=1}^{N} \mathbf{p}_i$)

Kinetic energy: $K(\{\mathbf{p}\}) = \sum_{i=1}^{N} \tfrac{1}{2} m_i |\mathbf{u}_i|^2 = \sum_i \frac{|\mathbf{p}_i|^2}{2m_i}$

($T$ = temperature , $K = \tfrac{3}{2} NkT$)

Potential energy: $\Phi(\{\mathbf{x}\})$ , $\mathbf{F}_i = -\frac{\partial \Phi}{\partial \mathbf{x}_i}$

($\mathbf{F}_i$ = force on particle $i$)

Phase-space point ($6N+1$ dimensions): $\Gamma = \{\mathbf{x}, \mathbf{p}, \xi\}$

Total energy: $E(\Gamma) = K(\{\mathbf{p}\}) + \Phi(\{\mathbf{x}\}) + E_\xi(\xi)$

$E_\xi = \tfrac{3}{2} NkT_0 \tau^2 \xi^2$ (thermostatting NH heat-bath energy)

We can now introduce the Nosé-Hoover equations of motion for thermostatting the system (either “sample” or “reservoir” in the Demonic system), where $\xi$ = thermostatting rate, $T_0$ = thermostat temperature, and $\tau$ = collision time:

$$\dot{\mathbf{x}}_i = \mathbf{u}_i + \dot{\mathbf{x}}_0 , \tag{1}$$

$$\dot{\mathbf{u}}_i = \frac{\mathbf{F}_i}{m_i} - \xi \mathbf{u}_i , \tag{2}$$

$$\dot{\xi} = \frac{1}{\tau^2}\left( \frac{T}{T_0} - 1 \right) . \tag{3}$$

We can then compute rate of change of the total energy, $dE/dt = dQ/dt - dW/dt$, of the NH thermostatted computational box, where $Q$ is the heat injected *into* the system from the thermostat heat bath, and $W$ is the work done *by* the system on its surroundings (e.g., work done to shear the fluid, or lift a weight, or change pressure or volume, etc.). From this, we can find the rate of change in the entropy, $S$, namely, $dS/dt$.

$$\dot{E} = \dot{K} + \dot{\Phi} + \dot{E}_\xi = \dot{Q} - \dot{W} \,, \quad \dot{Q} = T_0 \dot{S}$$

$$\dot{K} + \dot{\Phi} = \sum_i \left( m_i \mathbf{u}_i \cdot \dot{\mathbf{u}}_i + \frac{\partial \Phi}{\partial \mathbf{x}_i} \cdot \dot{\mathbf{x}}_i \right)$$

$$= \sum_i \left[ m_i \mathbf{u}_i \cdot \left( \frac{\mathbf{F}_i}{m_i} - \xi \mathbf{u}_i \right) - \mathbf{F}_i \cdot \mathbf{u}_i \right] \quad (\dot{\mathbf{x}}_0 \equiv 0)$$

$$= -\xi \sum_i m_i \,|\, \mathbf{u}_i \,|^2 = -3NkT\xi$$

$$\dot{E}_\xi = 3NkT_0 \tau^2 \xi \dot{\xi} = 3NkT_0 \xi \left( \frac{T}{T_0} - 1 \right) = 3NkT\xi - 3NkT_0 \xi$$

Thermostatting only, i.e., $\dot{W} \equiv 0$:

$$\dot{E} = \dot{Q} = -3NkT_0 \xi = T_0 \dot{S} \quad \Rightarrow \quad \dot{S} = -3Nk\xi \,. \qquad (4)$$

The NVT temperature fluctuation $\sigma_T / T_0 = [2/(3N)]^{1/2}$ is derived in Appendix A (as in Ref. [5]). The equations of motion (with initial CM velocity = 0) can be integrated by Størmer finite centered differences (as in Ref. [5]), where coordinates and the heat-flow variable are evaluated at integral values of the time step (as are forces, which depend on coordinates) while velocities are staggered by half time steps (as is the temperature). These centered difference equations of motion are time-reversible and avoid introducing center-of-mass drift errors, as well as systematic time-step errors (see details in Appendix B).

At equilibrium, long-time averages are equivalent to ensemble averages; for the nonequilibrium steady state (NESS), time averages are the whole game. For functions like temperature that fluctuate around a constant:

$$\left\langle T \right\rangle = \lim_{t \to \infty} \frac{1}{t} \int_0^\infty ds\, T(s) = T_0 \,,$$

$$\left\langle \dot{T} \right\rangle = \lim_{t \to \infty} \frac{1}{t} \int_0^\infty ds\, \dot{T}(s) = \lim_{t \to \infty} \frac{T(t) - T(0)}{t} = 0 \,.$$

For the Gaussian distribution of the thermostatting coefficient and its number-dependence at equilibrium (see Appendix A):

$$\langle \xi \rangle = 0\,, \quad \sqrt{\langle \xi^2 \rangle} = \frac{1}{\sqrt{3N} \cdot \tau}\,.$$

## II. MAXWELL DEMON: NONEQUILIBRIUM HEAT FLOW

Now, we are prepared to take these foundational concepts of NH thermostatting and generalize them to the two-state nonequilibrium heat-flow problem addressed by the Maxwell Demon [4]. NESS long-time averages of temperature, energy, and entropy are related to long-time averages of the two thermostat variables:

$$\begin{aligned}\langle \dot{E} \rangle &= -3Nk\langle T_0 \rangle \langle \xi_0 \rangle - 3Nk\langle T_1 \rangle \langle \xi_1 \rangle \\ &= -3NkT_0 \langle \xi_0 \rangle - 3NkT_1 \langle \xi_1 \rangle\end{aligned}$$

Nonequilibrium steady state (NESS): $\langle \dot{E} \rangle = 0$

$$\Rightarrow \quad \langle \xi_1 \rangle = -\frac{T_0}{T_1} \langle \xi_0 \rangle < 0 \qquad (5)$$

$$\begin{aligned}&\langle \dot{E} \rangle = \langle \dot{Q}_0 \rangle + \langle \dot{Q}_1 \rangle = T_0 \langle \dot{S}_0 \rangle + T_1 \langle \dot{S}_1 \rangle \\ &\langle \dot{S}_0 \rangle = -3Nk\langle \xi_0 \rangle\,, \quad \langle \dot{S}_1 \rangle = -3Nk\langle \xi_1 \rangle \\ &\therefore \quad \langle \dot{S} \rangle = \langle \dot{S}_0 \rangle + \langle \dot{S}_1 \rangle = -3Nk\left[ \langle \xi_0 \rangle + \langle \xi_1 \rangle \right] \\ &\qquad = -3Nk\langle \xi_0 \rangle \left( \frac{T_1 - T_0}{T_1} \right) < 0\,. \qquad (6)\end{aligned}$$

Clearly, the entropy never becomes steady, but drops forever as the phase-space distribution collapses onto a fractal strange attractor of lower dimensionality $< 6N+1$. (An alternative derivation of this nonequilibrium behavior of the entropy rate of change can be gotten from the Liouville continuity equation for the phase-space distribution function; see Appendix C.) The hot thermostat variable's long-time average is smaller in magnitude than the cold ($|\xi_1| < \xi_0$), by virtue of the conservation of energy, Eq.5. From Eq.6, we see that entropy continues to get more and more negative in steady-state heat flow, demonstrating conclusively that it is easier to heat a sample than to cool it.

The relationship of these NEMD heat-flow simulations to the Ibáñez lab experiments is clear, though the data from the NEMD thermostatting coefficients might have an analogue in steady-state laboratory experiments: if the experimental power required to keep the two temperature regions at their respective temperatures were measured, they might show a bias similar to the absolute value of the NEMD thermostatting coefficients.

The heat flow results of NEMD in a three-dimensional multi-particle NH thermostatted fluid can be compared to the much simpler, yet very instructive two-dimensional model of a single particle moving in a periodic cell, with temperature in the two directions thermostatted – one direction hot ($x$) at temperature $T_1 = T_x$ and one cold ($y$) at $T_0 = T_y$. The statistical mechanics analysis presented here for NEMD simulations in 3D ($N >> 1$) applies equally well to the much simpler 2D ($N = 1$). As we will now show, both give identical nonequilibrium predictions about entropy rate and ease of heating compared to cooling.

## III. EQUILIBRIUM SYMMETRY AND NONEQUILIBRIUM ASYMMETRY

### A. Single-Particle Cell Model in Two Dimensions

To investigate the statistical behavior of Nosé-Hoover thermostats under nonequilibrium conditions, we used the results of our recent study of a two-dimensional cell model with independent thermostats [3] applied along the $x$ and $y$ directions. The cell model potential energy is given by $\varphi(r) = \mathbf{\Sigma}_{i=1:4}\, \varepsilon[1 - |(\mathbf{r}/\sigma) - \boldsymbol{\delta}_i|^2\,]^4$ for $|(\mathbf{r}/\sigma) - \boldsymbol{\delta}_i| < 1$, otherwise, $\varphi(r) = 0$; the four repulsive corners (energy $\varepsilon$) of the PBC cell are $\boldsymbol{\delta}_i = (\pm 1, \pm 1)$ in reduced units: $m = \sigma = \varepsilon = k = 1$. The extended thermostat variables $\xi_x$ and $\xi_y$ are tracked in both equilibrium and NESS. Fig.1 shows the histograms of $\xi_x$ and $\xi_y$ for two cases, equilibrium: $T_x = T_y = 0.5$ ($\delta = T_x - T_y = 0$), and strongly nonequilibrium: $T_x = 0.5$, $T_y = 0.05$ ($\delta = 0.45$). In equilibrium, both thermostat variables are symmetrically distributed around zero with negligible skewness, as expected from time-reversible dynamics. Under nonequilibrium conditions, the distributions become highly asymmetric, with clear shifts in the mean and median values. These results are shown in Table I.

TABLE I: Mean and median values of $\xi_x$ and $\xi_y$ in equilibrium and NESS.

| Case | Variable | Mean | Median |
|---|---|---|---|
| Equilibrium (δ = 0.00) | $\xi_x$ | −0.00008 | −0.00011 |
| | $\xi_y$ | −0.00008 | −0.00011 |
| Nonequilibrium ($\delta$ = 0.45) | $\xi_x$ | −0.01541 | −0.02162 |
| | $\xi_y$ | +0.15415 | +0.14654 |

In NESS, we observe $\langle \xi_x \rangle < 0$ implies heating along $x$ and $\langle \xi_y \rangle > 0$ implies cooling along $y$. This reflects the imposed thermal gradient: heat flows from the hot direction ($x$) to the cold direction ($y$), and the thermostats respond accordingly. Interestingly, the thermostat effort is highly asymmetric, explained analytically by the balance equation for steady-state energy input and output, outlined in our foregoing Maxwell Demon scenario (Eq.5).

The thermostat along the cold direction ($y$) must apply an order-of-magnitude stronger damping force to absorb the incoming energy flux, while the hot direction ($x$) requires only mild anti-damping to inject energy. The results obtained here are in good agreement with both equilibrium and nonequilibrium cases of the many-body simulations as substantiated by the statistical mechanical arguments we present in this paper, summarized in Eq.5. Thus, the heating is easier to accomplish than cooling, confirming at the atomistic level the lab experimental results of Ibáñez et al. [1].

FIG. 1: Nosé-Hoover thermostat variable distributions in the two-temperature cell model [3]: for both (a) and (b) at equilibrium ($\delta$ = 0), P($\xi_x$) and P($\xi_y$) are centered around zero; for (c) and (d) under strongly nonequilibrium conditions ($\delta$ = 0.45), P($\xi_x$) and P($\xi_y$) develop strong asymmetry. The cold bath requires significantly stronger damping effort ($\xi_y >> |\xi_x|$) to maintain the steady-state heat flow. In reduced units, trajectory lengths were $t = 10^8$ (time-step 0.001), sampled once every integer time unit into histograms of bin size $\Delta\xi = 0.005$. (Note that the area of the finely spaced histograms have been shaded gray.)

**B. Equilibrium Asymmetry In Heating versus Cooling? Evans and Holian**

One of us (BLH) recalled, upon seeing the Ibáñez results, that Evans and Holian [6], in a 1985 molecular-dynamics (MD) computer simulation of a Nosé-Hoover (NH) thermostatted fluid, observed a similar asymmetry favoring heating – but at equilibrium! The asymmetry observed in the distribution of $\xi$, skewed toward negative values (favoring heating), were considered remarkable, since theory predicts a zero mean. Reflecting upon the Maxwell Demon hot-and-cold simulations and considering these striking findings of asymmetry at equilibrium, as well as thinking about the lessons learned from the simple cell model, where the equilibrium distribution of the thermostatting coefficient was perfectly symmetric and Gaussian, we thought it would be important to do a modern MD simulation, thermostatted at equilibrium by NH dynamics, for a fluid of $N = 32$ particles interacting by the purely repulsive Lennard-Jones soft-sphere potential, $\varphi(r) = 4\varepsilon(\sigma/r)^{12}$, with cut-off $r_c = 1.5\sigma$. (The full LJ potential has an additional attractive $r^{-6}$ term, well-depth $\varepsilon$, and zero energy at $r = \sigma$; LJ reduced units: $m = \sigma = \varepsilon = k = 1$.)

Why re-do this old work? We needed to assure ourselves that the earlier work did not have some kind of systematic error, either some kind of inaccuracy in the integration of the NH thermostatting equations of motion, or possible failure to maintain zero CM velocity, or the possible existence of finite system-size effects, or too short sampling times along the trajectory. Hence, the old and new system sizes and trajectory lengths, not to mention careful centered-difference integration of the NH thermostatting equations of motion, were deemed to be necessary for a definitive test of asymmetry – or *not* – in the thermostat variable's distribution at equilibrium.

Fig.2 shows the probability distributions P($\xi$) for all three cases. In the short trajectory that reproduces the original Evans and Holian setup, the newly calculated distribution is Gaussian and centered at $\langle\xi\rangle = 0$, in contrast to the skewed histogram reported in Fig.1 of Ref. [6]. We also confirmed that extending the trajectory length does not introduce any bias in $\langle\xi\rangle$. Both long simulations show symmetric, zero-mean Gaussian distributions, with predicted number dependence. Excellent agreement between the histograms and theoretical Gaussian confirms that the friction coefficient, $\xi$, behaves as a Gaussian random variable at equilibrium. (For a discussion of our two methods of error estimation in MD simulations, see Appendix D.)

FIG. 2: Probability distributions of thermostat variable P($\xi$) for three equilibrium simulations of a soft-sphere LJ fluid, where the Nosé-Hoover thermostat relaxation time $\tau$ = 0.09622: $N$ = 32 (trajectory lengths $t$ = 120 and $t$ = 120,000, time-step 0.003); $N$ = 1000 ($t$ = 120,000, time-step 0.001). To overlay $N$ = 1000 data onto $N$ = 32, $\xi$ is scaled by $\sqrt{(1000/32)} \approx 5.6$, justified by the Gaussian distribution of the variance (see Appendix A). All distributions are symmetric about zero and Gaussian in nature, in contrast to Ref. [6]; compare with Fig.1-a,b for cell model results at equilibrium. (Histograms have been shaded for the three simulations. Data points for the large system's long trajectory, for example, were collected once every ten collision times into bins of size $\Delta\xi$ = 0.1; for the small system's short trajectory, data was sampled every time-step.)

## IV. CONCLUSIONS

We have demonstrated at the microscopic scale, using Nosé-Hoover thermostatting in both many-body molecular-dynamics computer experiments and a very simple, but incredibly instructive single-particle two-dimensional cell model [3], that true asymmetry between heating and cooling arises not in equilibrium, but in driven nonequilibrium steady states.

The theory that allows us to quantify the rates was first outlined in earlier MD simulations of thermostatted side-by-side hot-and-cold regions [4]. By itself, this theory should have alerted us, two decades ago, to the inescapable fact of nonequilibrium heating/cooling asymmetry. The steady-state positive value of the thermostatting rate for the cold sample was reported; unfortunately, the paper gave no clue about the magnitude of the negative coefficient for pumping heat into the hot side – a missed opportunity!

In a 2D two-temperature one-particle cell model [4], NH thermostatted under nonequilibrium conditions, i.e., $T_x > T_y$, the mean thermostat variables become asymmetric, depending on the degree of deviation from linear response theory. This nonequilibrium asymmetry is explained analytically by a steady-state energy balance, leading to the relation (Eq.5): $\langle\xi_x\rangle/\langle\xi_y\rangle = -T_y/T_x$.

Prompted by lab experiments showing asymmetric heating and cooling rates [1], we revisited and resolved a discrepancy first reported by Evans and Holian [6] in the statistical behavior of the NH thermostat at equilibrium. While their original simulation appeared to show a skewed distribution for the thermostat variable $\xi$, we carried out modern simulations under identical initial conditions, but including strict conservation of center-of-mass momentum and the long-time average of internal energy by careful time-reversible finite-difference integration of the equations of motion. The results reveal that the thermostat rate variable exhibits a totally symmetric zero-mean Gaussian distribution, in agreement with equilibrium theory and the cell model, and validated by using two independent error estimation methods (see Appendix D). We emphasize that symmetry in heating and cooling is an *equilibrium* property only.

These findings clarify the nonequilibrium conditions under which thermostat asymmetry emerges and offer a microscopic explanation for recent experimental observations of unequal heating and cooling rates [1]. We have shown that Nosé-Hoover friction variables not only regulate energy exchange but also encode the thermodynamic structure of nonequilibrium states, including the chaotic phase-space contraction to a lower-dimensional, fractal, strange attractor. Moreover, our computer experiments have demonstrated relevance to real-world lab experiments.

As a wise old cowboy once said, “When you’re riding your horse, always be prepared for the unexpected.” We observe that the same applies to nonequilibrium systems.

## APPENDIX A. FLUCTUATIONS UNDER NOSÉ-HOOVER THERMOSTATTING

In this Appendix, we derive thermal fluctuations under Nosé-Hoover (NH) thermostatting. The NVT equilibrium canonical ensemble ($N$ particles in volume $V$ and temperature $T$) is characterized by the phase-space distribution function $\rho(\mathbf{\Gamma})$ at the $6N$+1-dimensional phase point, $\mathbf{\Gamma} = \{\mathbf{x},\mathbf{p}, \xi\}$, which is the collection of $3N$ coordinates $\mathbf{x}$, $3N$ momenta $\mathbf{p}$, and the NH thermostat heat-flow variable $\xi$: $\rho(\mathbf{\Gamma}) = \exp(-\beta E)/Q_{\mathrm{NVT}}$, where $\beta=1/kT_0$, $k$ is Boltzmann's constant, $T_0$ is the canonical-ensemble temperature set by the NH thermostat, $E = K + \Phi + E_\xi$ is the total energy in the exponential Boltzmann factor – the usual internal energy, kinetic $K$ and potential $\Phi$ – plus a harmonic thermostat energy, $E_\xi = (3/2)NkT_0\tau^2\xi^2$, representing the NH heat bath [5], and $Q_{\mathrm{NVT}}$ is the canonical partition function, the integral of the Boltzmann factor over all of phase space. Since $\rho(\mathbf{\Gamma}) = 1/\Omega$, where $\Omega$ is the number of microstates for the system at $\mathbf{\Gamma}$, the entropy of the system is $S = -k \ln \rho = k \ln \Omega$ (Boltzmann's epitaph on his tomb in Vienna).

We can define the dimensionless thermostat strength, characterizing the exchange of energy between the NH heat bath and the $N$ sample particles, as $\zeta = \xi\tau$, where $\xi$ is the thermostatting rate variable and $\tau$ is the thermostat relaxation time (most efficiently chosen as the collision time). Then, the harmonic heat-bath "potential" is $E_\xi = (1/2)\kappa\zeta^2$, where $\kappa = 3NkT_0$ is the "force constant" or "stiffness" of the thermostatting potential well. If either $\zeta > 0$ (so that friction takes kinetic energy out of the $3N$ particle momenta by slowing them down) or if $\zeta < 0$ (so that anti-friction pumps energy into the momenta, speeding them up) – in either case – there is an energetic penalty for thermostatting added to the $6N$+1–dimensional total energy of system. The phase-space distribution function contribution from the NH heat bath $\zeta$ ends up being a simple Gaussian: $\exp(-\beta E_\xi) = \exp(-3N\zeta^2/2)$, whereby the variance of the dimensionless thermostatting strength is $\langle\zeta^2\rangle = 1/(3N)$. Hence, the ensemble average (or time average) is $\langle E_\xi\rangle = (1/2)\kappa\langle\zeta^2\rangle = kT_0/2$, which is a factor of $3N$ smaller than the particle momentum contribution to the kinetic energy: $\langle K\rangle = 3N(kT_0/2)$. For a monatomic harmonic crystalline solid, $\langle \Phi\rangle = \langle K\rangle$, so that each of the $3N$ normal modes has a contribution from the particle coordinates of $kT_0/2$, equal to that of the momenta; for the ideal gas, $\Phi$ = 0; monatomic fluids fall in between these two idealized limits.

The whole idea of thermostatting is to approach the NVT thermodynamic limit long before $N$ becomes infinite, at which point, the NH thermostat contributes nothing to the system.

Finally, the temperature (thermal) fluctuation, $\sigma_T^2$, is obtained by noticing that the NVT phase space can be separated into independent partition functions governing **x**, **p**, and $\xi$, and evaluating ensemble averages within them.

$$E = K(\mathbf{p}) + \Phi(\mathbf{x}) + E_\xi(\xi)$$

Independent phase spaces:

$$Q_{\mathrm{NVT}}(\Gamma) = \int d\Gamma \exp(-\beta E) = Q_K(\mathbf{p}) Q_\Phi(\mathbf{x}) Q_\xi(\xi),$$

$$\text{e.g.,} \quad Q_K = \int d\mathbf{p} \exp(-\beta K)$$

$$\langle K \rangle = \frac{3}{2} Nk \langle T \rangle = \frac{3}{2} NkT_0 = \frac{1}{Q_K} \int d\mathbf{p}\; K \exp(-\beta K)$$

$$= -\frac{1}{Q_K} \frac{\partial}{\partial \beta} \int d\mathbf{p} \exp(-\beta K) = -\frac{1}{Q_K} \frac{\partial Q_K}{\partial \beta}$$

$$\langle K^2 \rangle = \frac{1}{Q_K} \int d\mathbf{p}\; K^2 \exp(-\beta K)$$

$$= -\frac{1}{Q_K} \frac{\partial}{\partial \beta} \int d\mathbf{p}\; K \exp(-\beta K)$$

$$= -\frac{1}{Q_K} \frac{\partial Q_K}{\partial \beta} \cdot \langle K \rangle - \frac{\partial \langle K \rangle}{\partial \beta}$$

$$= \langle K \rangle^2 + kT_0^2 \frac{\partial \langle K \rangle}{\partial T_0}$$

$$= \langle K \rangle^2 + \frac{3}{2} Nk^2 T_0^2$$

$$\langle K^2 \rangle - \langle K \rangle^2 = \frac{3}{2} Nk^2 T_0^2 = \left( \frac{3}{2} Nk \right)^2 \left( \langle T^2 \rangle - \langle T \rangle^2 \right)$$

$$\Rightarrow \quad \sigma_T^2 = \langle T^2 \rangle - \langle T \rangle^2 = \frac{2T_0^2}{3N} \qquad \text{(A-1)}$$

Fluctuation in temperature vanishes in the thermodynamic limit.

## APPENDIX B. NOSÉ-HOOVER EQUATIONS OF MOTION ON THE COMPUTER

Nosé-Hoover equations of motion, Eqs.1-3 (with initial CM velocity = 0) can be integrated by Størmer finite centered differences [5], where coordinates and the heat-flow variable are evaluated at integral values of the time step $\delta$ (as are forces, which depend on coordinates) while velocities are staggered by $\delta/2$ (as is the temperature). Suppressing particle indices $i = 1{:}N$, and vector notation $\alpha = 1{:}3$ in 3D,

$$x(t) = x(t-\delta) + u(t-\tfrac{1}{2}\delta)\cdot\delta + O(\delta^3) \qquad \text{(B-1)}$$

$$\xi(t) = \xi(t-\delta) + \frac{1}{\tau^2}\left[T(t-\tfrac{1}{2}\delta)/T_0 - 1\right]\cdot\delta + O(\delta^3)\,, \qquad \text{(B-2)}$$

$$3NkT(t-\tfrac{1}{2}\delta) = \sum_i\sum_\alpha m_i u_{i\alpha}^2(t-\tfrac{1}{2}\delta)$$

$$u(t+\tfrac{1}{2}\delta) = \frac{u(t-\tfrac{1}{2}\delta)\left[1-\tfrac{1}{2}\xi(t)\cdot\delta\right] + a(t)\cdot\delta}{1+\tfrac{1}{2}\xi(t)\cdot\delta} + O(\delta^3)\,, \qquad \text{(B-3)}$$

$$a(t) = F[x(t)]/m \quad \text{(Newton: } F = ma\text{)}$$

The local error in the coordinate, velocity, and thermostat variable equations of motion is third-order, i.e., the Taylor series terms in $\delta^3$ and higher have been excluded in the finite difference equations above. The two-step error, $\delta^4$, is obtained by integrating the coordinate one more time (to $t + \delta$). The global error in central difference integration after $n$ steps ($t_n = n\delta$) is $O(\delta^2)$, i.e., second order:

$$n=1: \quad x_1 = x_0 + \dot{x}_0\delta + \tfrac{1}{2}\ddot{x}_0\delta^2\,,$$

$$\text{exact solution}: \quad q_1 = x_0 + \dot{x}_0\delta + \tfrac{1}{2}\ddot{x}_0\delta^2 + \tfrac{1}{6}x_0^{\mathrm{III}}\delta^3 + \cdots$$

$$\frac{x_1}{q_1} = 1 - \tfrac{1}{6}\frac{x_0^{\mathrm{III}}}{x_0}\delta^3 + \cdots$$

$$\text{time } t_n: \quad \frac{x_n}{q_n} = \left(\frac{x_1}{q_1}\right)^n = \left(1 - \tfrac{1}{6}\frac{x_0^{\mathrm{III}}}{x_0}\delta^3 + \cdots\right)^n$$

$$= 1 - \frac{n\delta}{6}\frac{x_0^{\mathrm{III}}}{x_0}\delta^2 + \cdots$$

$$= 1 - \frac{t_n}{6}\frac{x_0^{\mathrm{III}}}{x_0}\delta^2 + \cdots \qquad \text{(B-4)}$$

An important property of finite-difference equations of motion is time-reversibility, which leads to long-term stability of the trajectory, particularly in keeping the long-time average of the internal energy constant.

Time-reversibility under finite central difference integration:

$$\bullet \text{———}\!\rightarrow \bullet \text{———}\!\rightarrow \bullet$$

$$x(t-\delta) \qquad x(t) \qquad x(t+\delta) \qquad (\textit{coordinate})$$

$$u(t)\longrightarrow \qquad (\textit{velocity})$$

$$\rightarrow a(t) \qquad (\textit{acceleration})$$

time $= 0 \rightarrow t$, reverse velocities,

then return to initial conditions, $t' = t \rightarrow 0$ :

$$\bullet\!\leftarrow\!\text{———} \bullet\!\leftarrow\!\text{———} \bullet$$

$$x'(t'+\delta) \quad x'(t') \quad x'(t'-\delta)$$

$$\longleftarrow u'(t') = -u(t)$$

$$\rightarrow a'(t') = a(t)$$

$$x'(t') = x(t)\,,\; x'(t'+\delta) = x(t-\delta)\,,\; x'(t'-\delta) = x(t+\delta)$$

$$u'(t') = \frac{x'(t'+\delta) - x'(t'-\delta)}{2\delta} + O(\delta^2)$$

$$= -\frac{x(t+\delta) - x(t-\delta)}{2\delta} + O(\delta^2) = -u(t) \qquad \text{(B-5)}$$

$$a'[x'(t')] = a[x(t)] = F[x(t)]/m$$

NH thermostat rate behaves like velocity, $\dot{Q} = -3NkT_0\xi$ :

$$\xi'(t') = -\xi(t) \qquad \text{(B-6)}$$

$\dot{\xi}$ behaves like acceleration : $\dot{\xi}'[\mathrm{u}'^2(t')] = \dot{\xi}[u^2(t)] = \dot{\xi}[T(t)]$

$$\dot{u}'(t') = a'(t') - \xi'(t')u'(t')$$

$$= \frac{x'(t'+\delta) - 2x'(t') + x'(t'-\delta)}{\delta^2} + O(\delta^2)$$

$$= \frac{x(t-\delta) - 2x(t) + x(t+\delta)}{\delta^2} + O(\delta^2)$$

$$= a(t) - \xi(t)u(t) = \dot{u}(t) \qquad \text{(B-7)}$$

## APPENDIX C. LIOUVILLE DERIVATION: NONEQUILIBRIUM ENTROPY CHANGE

The Liouville continuity equation applied to the phase-space distribution function $\rho(\mathbf{\Gamma},t)$ at the $6N$+1-dimensional phase point, $\mathbf{\Gamma} = \{\mathbf{x},\mathbf{p}, \xi\}$, and time $t$, is:

$$\frac{\partial \rho}{\partial t} + \frac{\partial}{\partial \Gamma} \cdot \rho \dot{\Gamma} = 0 = \frac{d\rho}{dt} + \rho \Lambda \,,$$

$$\frac{d\rho}{dt} = \frac{\partial \rho}{\partial t} + \frac{\partial \rho}{\partial \Gamma} \cdot \dot{\Gamma} \,,$$

$$\Lambda = \frac{\partial}{\partial \Gamma} \cdot \dot{\Gamma} = -\frac{d}{dt} \ln \rho$$

$$\Lambda = \sum_{\alpha} \sum_{i} \left( \frac{\partial \dot{x}_{i\alpha}}{\partial x_{i\alpha}} + \frac{\partial \dot{p}_{i\alpha}}{\partial p_{i\alpha}} \right) + \frac{\partial \dot{\xi}_0}{\partial \xi_0} + \frac{\partial \dot{\xi}_1}{\partial \xi_1}$$

$$= \sum_{\alpha} \sum_{i} \left( \frac{\partial \dot{p}_{i\alpha}}{\partial p_{i\alpha}} \right) = -3N\left(\xi_0 + \xi_1\right)$$

$$\Rightarrow \quad \left\langle \dot{S} \right\rangle = -k \frac{d}{dt} \ln \rho = k \left\langle \Lambda \right\rangle = -3Nk \left[ \left\langle \xi_0 \right\rangle + \left\langle \xi_1 \right\rangle \right] \qquad \text{(C-1)}$$

## APPENDIX D. ERROR ANALYSIS

*a. Standard Error of the Mean*

The standard error of the mean (e.g., for the thermostat variable) is estimated from $\sigma_{mean} = \sigma/\sqrt{N_{eff}}$, where $\sigma$ is the sample standard deviation, or square root of the variance, and $N_{eff}$ is the effective number of statistically independent samples along a dynamical trajectory. In a time series with persistence over a characteristic timescale, $\tau$ (the mean correlation time, or approximately the collision time $\tau_{collision}$, which, coincidentally, is the best choice for the thermostat relaxation time [5]), the number of samples is $N_{eff} \approx t/\tau$. This yields the standard error formula, $\sigma_{mean} \approx \sigma/\sqrt{(t/\tau)}$, given that $\tau \approx \tau_{collision}$ from the equation of state.

*b. Kinetic-Crossing Method*

An alternative to estimating the standard error of the mean is the *kinetic-crossing method* (see Fig. 2 of [6]). By counting the number of times the instantaneous kinetic temperature, $T(t)$, crosses its mean value, $\langle T \rangle$ (= the thermostat value, $T_0$), the method provides an estimate of the number of statistically independent fluctuations in a Nosé-Hoover thermostatted trajectory. Hence, $n_{cross} = t/\tau$ ( $\approx N_{eff}$ ), so that the standard error of the mean can be estimated as $\sigma_{mean} = \sigma/\sqrt{n_{cross}} \approx \sigma/\sqrt{(t/\tau)}$, where $\sigma$ is the standard deviation of the temperature over the trajectory. This method is simple to implement, requires no auto-correlation or fitting analysis from independent calculations of the local equation of state, and can be applied to any observable (e.g., energy, pressure, thermostat variables) using the same set of thermal zero-crossings. The difference between the $\tau$-based standard-error method and the kinetic-crossing approach is that the latter does an on-the-fly evaluation of $\tau \approx \tau_{collision}$.

We applied kinetic crossing to estimate the uncertainty in the NH friction coefficient, $\xi$, and compared it to the $\tau$-based estimate. In Table D-I, we summarize results for all three systems, $N$ = 32 short and long runs, and $N$ = 1000. The two error estimates show good agreement, indicating that the equation of state estimate of the collision time (= the thermostat relaxation time) was close to the correlation time obtained by the number of kinetic crossings.

The results of our MD simulations of a many-body soft-sphere fluid at equilibrium, far from the melting line, were definitive: the distribution of the thermostatting coefficient is almost perfectly Gaussian, with zero mean and median, and with the theoretical thermal width. We conclude that skewed distribution of $\xi$ at equilibrium reported in Ref. [6] is due to integration error in those simulations.

**TABLE D-I**: Comparison of standard error of mean estimates of $\xi$ using (1) kinetic-crossing for $N = 32$ systems, and (2) $\tau$-based methods for all systems.

| $N$ | $t$ | $\delta$ | $\sigma(n_{\mathrm{cross}})$ | $\sigma(t/\tau)$ |
|---|---|---|---|---|
| 32 | 120 | 0.003 | 0.0031 | 0.0028 |
| 32 | $1.2 \times 10^6$ | 0.003 | 0.00003 | 0.00003 |
| 1000 | $10^5$ | 0.001 | – | 0.00002 |

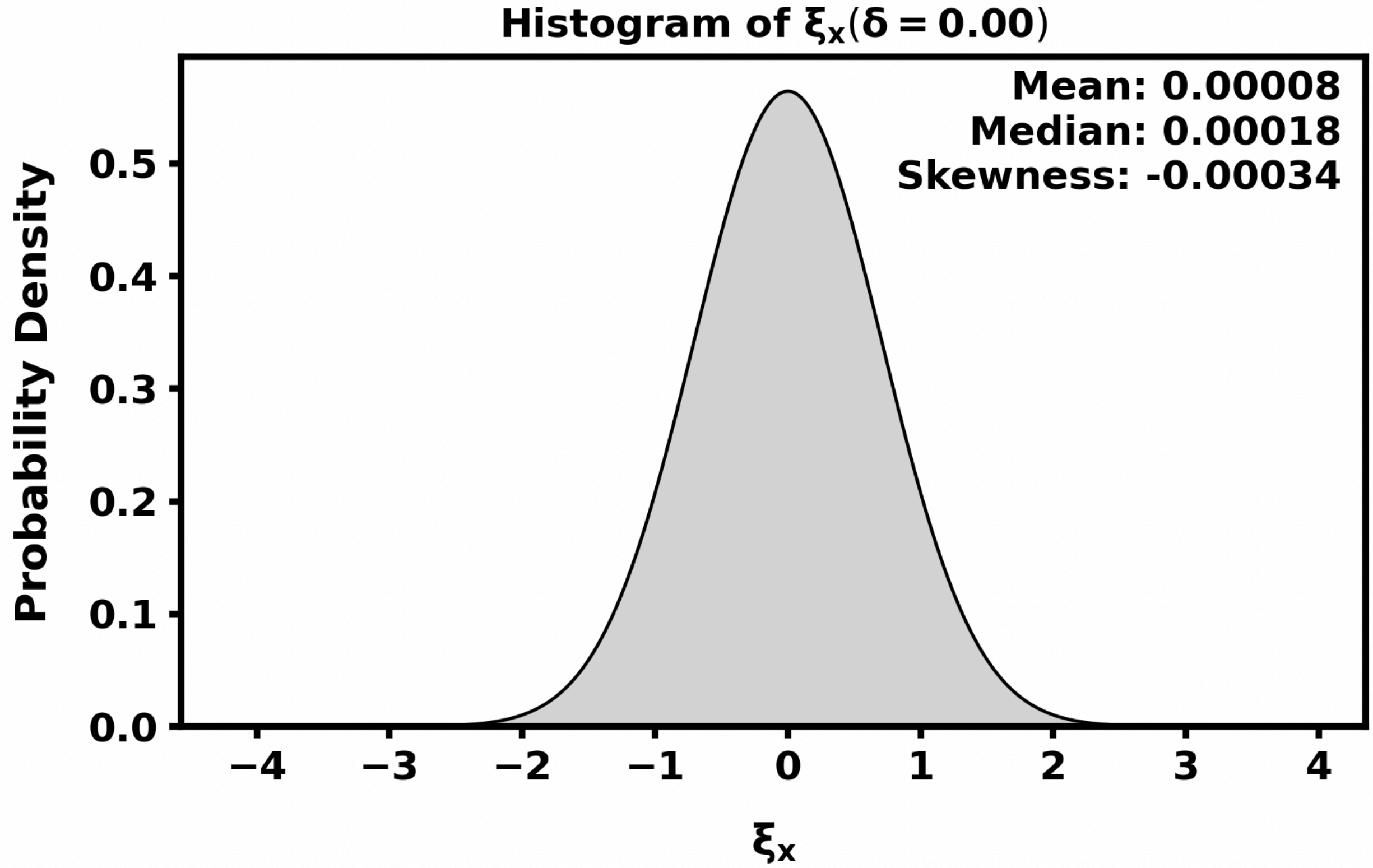

**FIG. 1 (a)**

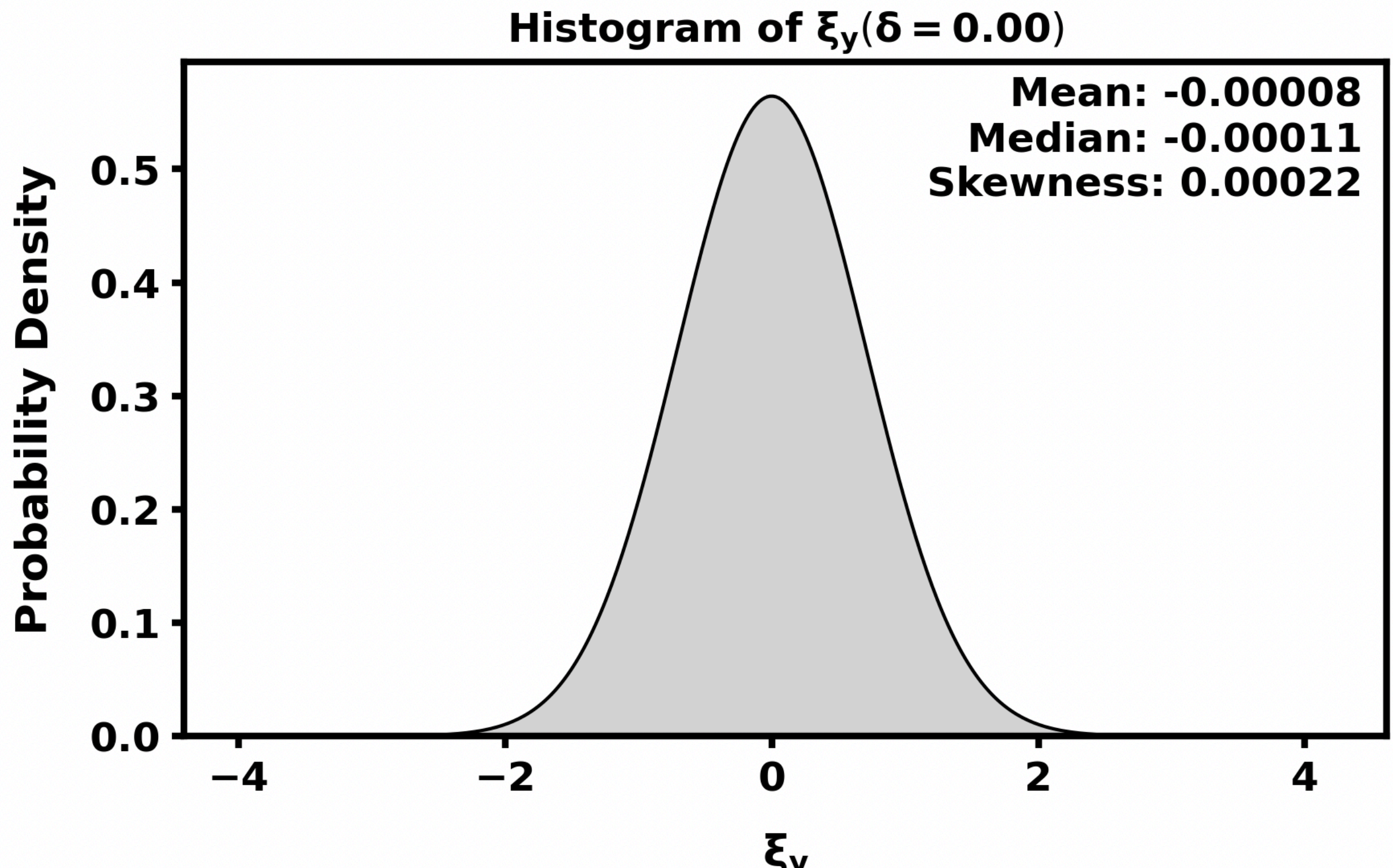

**FIG. 1 (b)**

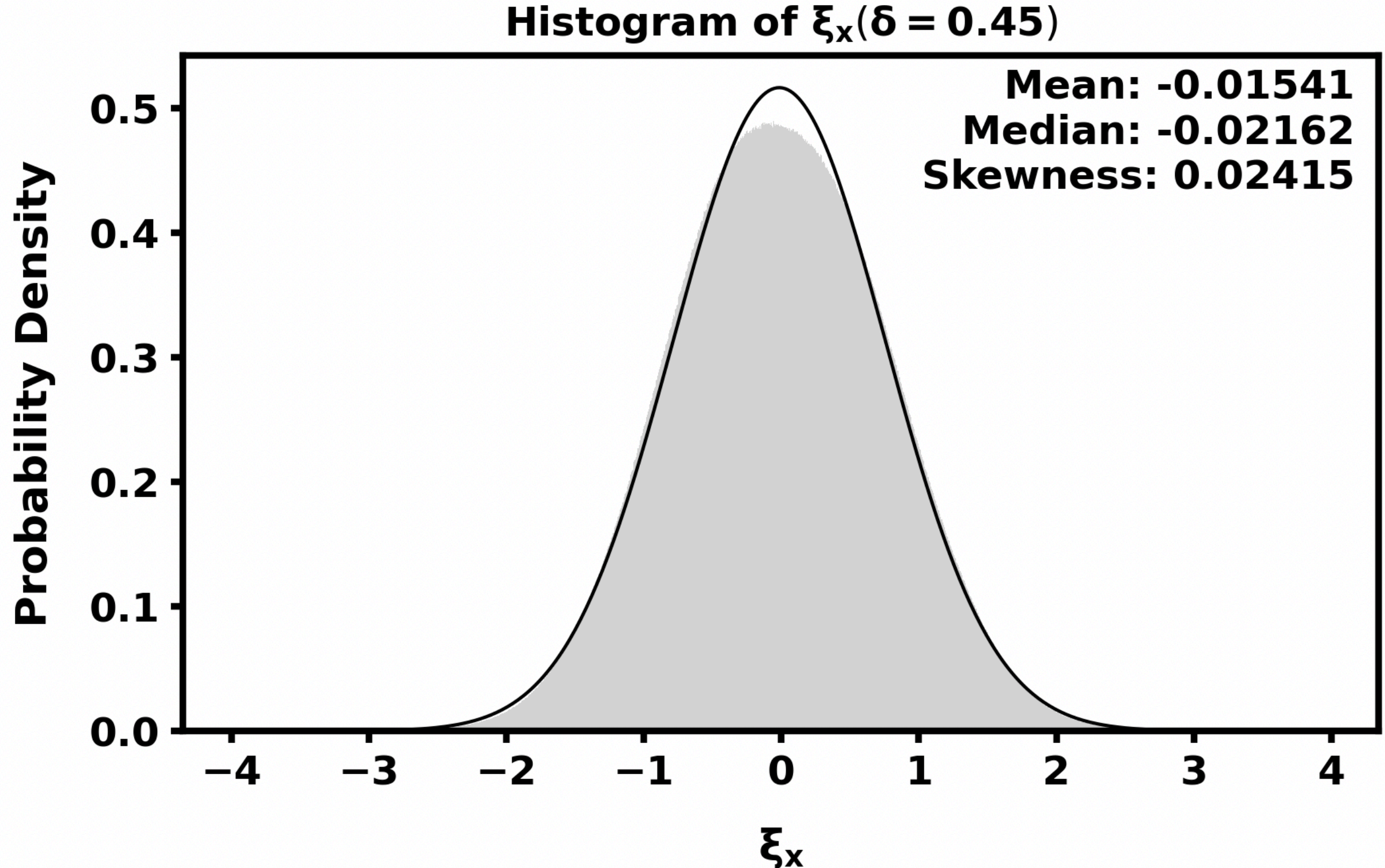

**FIG. 1 (c)**

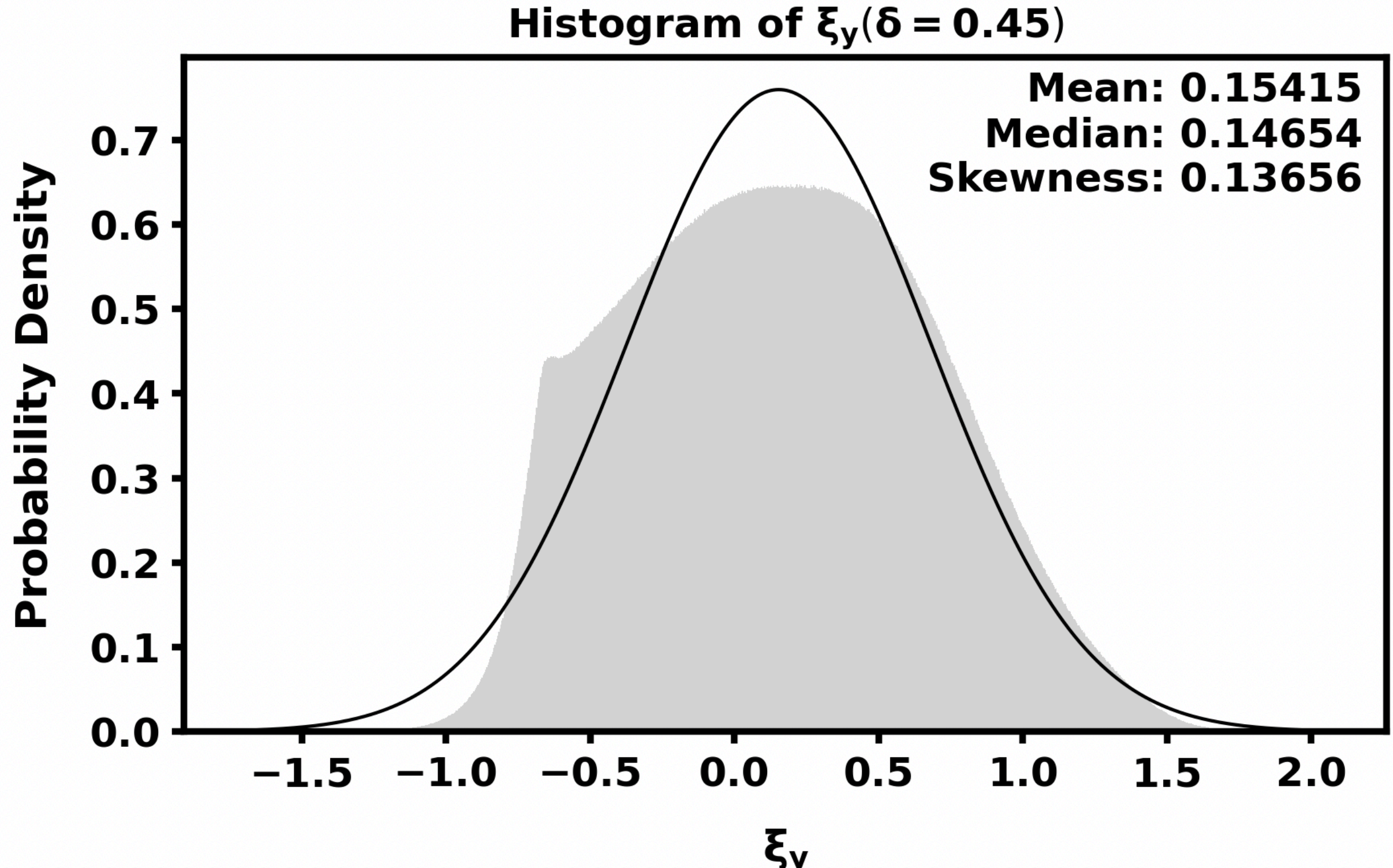

**FIG. 1 (d)**

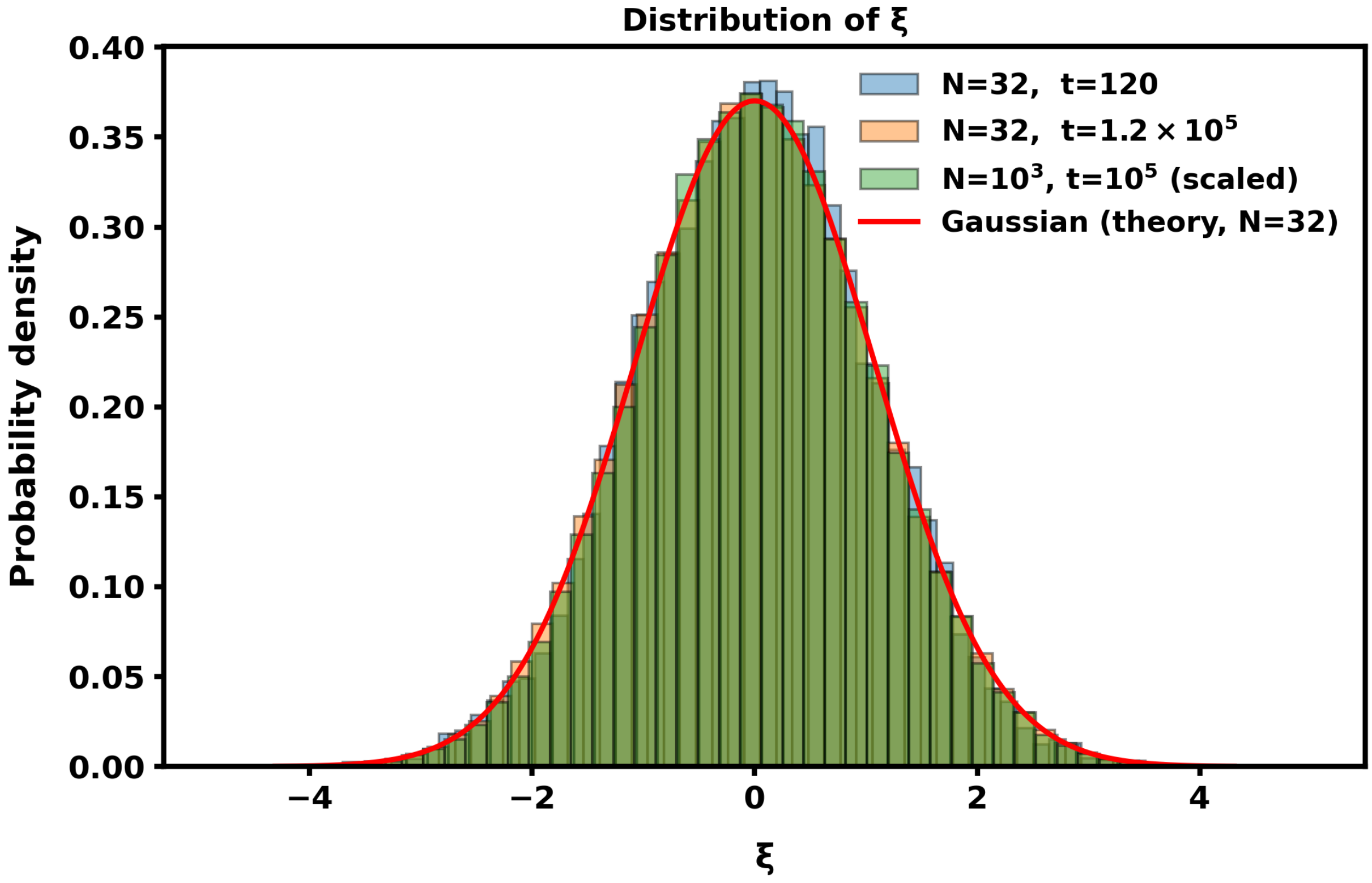

FIG. 2